\documentclass[aps,prb,twocolumn,psfig,showpacs]{revtex4-1}
\usepackage{amsfonts}
\usepackage{amsmath}
\usepackage[dvips]{graphicx}
\usepackage{bm}
\usepackage{amssymb}
\usepackage{times}
\usepackage{dcolumn}
\usepackage{cases}
\usepackage{txfonts}
\usepackage{color}

\RequirePackage[hyperindex,colorlinks,bookmarksnumbered,plainpages=true]{hyperref}
\hypersetup{colorlinks,linkcolor=blue,urlcolor=blue,citecolor=blue}
\usepackage{hyperref}

\usepackage{ulem}
\renewcommand{\emph}[1]{\textit{#1}}

\definecolor{darkgreen}{rgb}{0,0.5,0}
\definecolor{darkblue}{rgb}{0,0,0.5}
\definecolor{darkred}{rgb}{.7,0,0}
\definecolor{purple}{rgb}{0.35,0,0.35}
\definecolor{orange}{rgb}{1,0.5,0}
\definecolor{grey}{rgb}{.6,.6,.6}

\newcommand{\Eq}[1]{Eq.~(\ref{#1})}

\newcommand{\Fig}[1]{Fig.~\ref{#1}}

\begin{document}
\title{Featureless quantum spin liquid, 1/3-magnetization plateau state and exotic thermodynamic properties of spin-1/2 frustrated Heisenberg antiferromagnet on an infinite Husimi lattice}
\author{Tao Liu$^1$, Shi-Ju Ran$^1$, Wei Li$^{2,1}$, Xin Yan$^1$, Yang Zhao$^1$, and Gang Su$^1$}
\email{gsu@ucas.ac.cn}
\affiliation{$^{1}$Theoretical Condensed Matter Physics and Computational Materials Physics Laboratory, School of Physics, University of Chinese Academy of Science, Beijing 100049, China
\linebreak $^{2}$Physics Department, Arnold Sommerfeld Center for Theoretical Physics, and Center for NanoScience, Ludwig-Maximilians-Universit\"at, 80333 Munich, Germany}

\begin{abstract}
 By utilizing the tensor-network-based methods, we investigate the zero- and finite-temperature properties of the spin-1/2 Heisenberg antiferromagnetic (HAF) model on an infinite Husimi lattice that contains 3/2 sites per triangle. The ground state of this model is found to possess vanishing local magnetization and is featureless; the spin-spin and dimer-dimer correlation functions are verified to decay exponentially; and its ground-state energy per site  is determined to be $e_0=-0.4343(1)$, which is very close to that [$e_0=-0.4386(5)$] of the intriguing kagome HAF model. The magnetization curve shows the absence of a zero-magnetization plateau, implying a gapless excitation. A $1/3$-magnetization plateau with spin up-up-down state is observed, which is selected and stabilized by quantum fluctuations. A ground state phase diagram under magnetic fields is  presented. Moreover, both magnetic susceptibility and the specific heat are studied, whose low-temperature behaviors reinforce the conclusion that the HAF model on the infinite Husimi lattice owns a gapless and featureless spin liquid ground state.
\end{abstract}

\maketitle

\section{Introduction}

Spin liquids are disordered states in correlated spin systems, in which strong quantum fluctuations prevent from the formation of conventional spin orders related with spontaneous symmetry breaking.\cite{Balents, Yan} Such spin liquid states are common in one-dimensional (1D) systems owing to the low coordinate number and strong quantum fluctuations. However, for the systems beyond 1D, the correlated spins favor to freeze into solid at low temperatures. The two- or three-dimensional correlated spin systems that retain in spin liquid states at zero temperature are usually believed to be exotic quantum states, which are yet scarce in realistic, such as Heisenberg models. The first proposal of a quantum spin liquid (QSL) can be dated back to Anderson's resonant valence bond (RVB) ansatz for a triangular antiferromagnet.\cite{Anderson} The RVB wave function is a linear superposition of all possible valence bond configurations with some specified weights. The short-range RVB state is usually a nonmagnetic state, while some long-range RVB states can own N\'eel long-range order on a bipartite lattice.\cite{Liang} Very recently, it was revealed that the short-range RVB state is a gapped $Z_2$ spin liquid on the kagome lattice,\cite{Poilblanc} and is a gapless spin liquid on the $J_1$-$J_2$ square lattice.\cite{LWang}

\begin{figure}
\includegraphics[width=0.85\linewidth,clip]{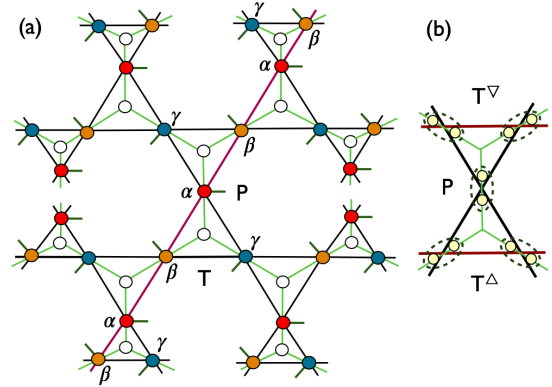}
\caption{(Color online) (a) Infinite Husimi lattice and corresponding tensor-network representation. Husimi lattice is composed of corner-sharing triangles, and the underlying tensor network structure is on a Bethe lattice. The empty circle represents a rank-3 tensor $T$ siting in the center of a triangle, and the colorful circles (red, blue and yellow for three sublattices $\alpha$, $\beta$ and $\gamma$, respectively) are rank-3 tensors $P$ with physical indices locating at the sharing vertices of triangles. (b) The underneath ancilla structures of tensors, the $T$ tensors ($T^{\triangle}$ and $T^{\triangledown}$ for up and down triangles, respectively) connects three ancillas in the same triangle, and the $P$ tensor (dashed oval) on the vertex projects two ancillas into physical space.}
\label{fig1}
\end{figure}

For a long time it has been thought that the spin liquids could be found in geometrically frustrated magnets.\cite{Lee} In general, geometric frustration leads to a huge degeneracy in classical spin configurations and enhances the fluctuations, thus a QSL ground state might be favored. However, the ground state of the spin-1/2 quantum HAF model on the triangular lattice was revealed to be an antiferromagnet with $120^{\circ}$ coplanar N\'eel order,\cite{Huse, Lhuillier} not a RVB spin liquid. Another highly frustrated lattice -- the intriguing kagome lattice with less coordinate number than the triangular lattice, has attracted much attention, both theoretically and experimentally, in the past decades.\cite{Sachdev, Leung, FWang, Ran, Hermele, Jiang1, Depenbrock, Jiang2,Didier, Shores, Mendels, Helton,Han} The HAF model on the kagome lattice is currently regarded as a very competitive candidate supporting the QSL ground state. A recent density matrix renormalization group (DMRG) simulation shows that the ground state of the kagome HAF model is a $Z_2$ spin liquid;\cite{Yan, Depenbrock, Jiang2} while some variational Monte Carlo simulations suggest a gapless U(1) Dirac spin liquid.\cite{Ran, Didier} This issue is still inconclusive. Theoretical predictions spurred experimentalists to search for a QSL in, for instance, mineral Herbersmithite. No signal of any magnetic order was observed down to 50 mK, that is much lower than the estimated Curie-Weiss temperature.\cite{Shores, Mendels, Helton} Fractional excitations have also been detected in Herbersmithite single crystal lately.\cite{Han} These studies give possible experimental evidences of a QSL, and some of them suggest that the low-energy excitations are gapless.\cite{Han, Vries}

In this paper, we study a spin-1/2 frustrated HAF model on the infinite Husimi lattice [\Fig{fig1} (a)], whose ground state is verified to be a gapless QSL. The infinite Husimi lattice, free of boundaries, consists of corner-sharing triangles, and has 3/2 sites per triangle (the same local structure as the kagome lattice). The HAF model on this infinite lattice is highly frustrated, which is very important for searching exotic spin-liquid states. It is worthy to mention that the finite-size Husimi cactus, which has roughly as many sites on the boundary as in the bulk, can be exactly solved because the HAF model in this case is frustration-free,\cite{explain-Husimi} and no spin liquid solution on Husimi cactus was found owing to a heavy boundary effect.\cite{explain-CTBL} Here we utilize the iterative approaches, i.e, tensor-network-based numerical simulations, to study the ground-state and thermodynamic properties of the HAF model with high accuracy. Our results provide ample and solid evidences manifesting the existence of a featureless QSL on the Husimi lattice.

This paper is organized as follows. The model and the adopted tensor-network approach are introduced in Sec. \uppercase\expandafter{\romannumeral2}. The quantum spin liquid ground state is discussed in Sec. \uppercase\expandafter{\romannumeral3}, and the magnetization curve and phase diagram are addressed in Sec. \uppercase\expandafter{\romannumeral4}. The thermodynamic properties of both susceptibility and specific heat are presented and discussed in \uppercase\expandafter{\romannumeral5}. Finally, Sec. \uppercase\expandafter{\romannumeral6} is devoted to the discussion and conclusion.

\begin{figure}
\includegraphics[width=1\linewidth,clip]{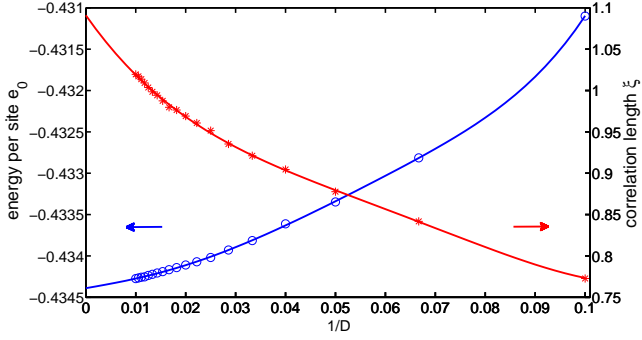}
\caption{(Color online) The calculated energy per site $e_0$ and the correlation length $\xi$ versus inverse bond dimension $1/D$ of the tree tensor network (up to $D=100$) for $\delta=1$. It is shown that $e_0$ decreases and $\xi$ increases with enhancing $D$, and $e_0$ converges faster than $\xi$. The solid lines are polynomial fittings, where $e_0$ is extrapolated to be -0.4344(1) in the infinite $D$ limit.}
\label{fig2}
\end{figure}

\section{Model and Method}
Let us consider the spin-1/2 XXZ HAF model on the infinite Husimi lattice, whose Hamiltonian reads
\begin{equation}\label{eq1}
  H = J \sum\limits_{<ij>}[(S^{x}_{i}S^{x}_{j}+S^{y}_{i}S^{y}_{j}) + \delta S^{z}_{i}S^{z}_{j}]-h\sum\limits_{i}S^{z}_{i},
\end{equation}
where $\textbf{S}_i$ is the spin operator on the $i$-th site, $J=1$ is the coupling constant which is set as the energy scale, $\delta$ is the XXZ anisotropy, and $h$ is the magnetic field. We use the tree tensor networks (TTNs) to simulate both the ground state and thermodynamic properties of the present model. The TTN is composed of simplex tensors $T$ and vertex tensors $P$ (see \Fig{fig1}), $T$-tensor sits on each triangle, while $P$-tensor with one (for the ground state) or two (for the thermal states) physical indices locates at the sharing vertex of two triangles. Note that the short-range RVB state on the kagome lattice is found to have an exact projected entangled-pair state (PEPS) representation with similar local tensor structures and the bond dimension is D=3.\cite{Poilblanc} Next, such kind of tensor structure was generalized to other states with larger bond dimensions.\cite{Xie}

In Fig. \ref{fig1}(a), the underneath tensor network forms a Bethe lattice. Owing to the loop-free structure, this Bethe-lattice TTN state can be processed easily and accurately by a simple update scheme during the imaginary-time evolution,\cite{Jiang3} which provides optimal truncations.\cite{Li, Liu} The Bethe (Husimi) lattice, although seemingly exists only as an ideal lattice, can actually be used to simulate the inner part of a large Cayley tree structure which can be synthesized in the lab (like dendrimers).\cite{Astruc} In the following, we introduce the ground-state projection algorithm for the infinite Husimi lattice. The imaginary time evolution for the finite-temperature thermal state, or equivalently the linearized tensor renormalization group (LTRG) process,\cite{LTRG, Ran_Su} can be implemented similarly.

On the Husimi lattice, there exist two kinds of $T$-tensors ($T^{\triangle, \triangledown}$) that correspond to up and down triangles, respectively. In order to implement the projections, we can group the three $P$-tensors, as well as the positive semi-definite diagonal matrices $\lambda$'s living on the virtual bonds linking tensors $T$ and $P$, with one up-triangle tensor $T^{\triangle}$ in odd steps (or down-triangle $T^{\triangledown}$ in even steps):
\begin{equation}
M^{m_1, m_2, m_3}_{x, y, z} = \sum_{x', y', z'=1}^{D} (\lambda_{1})_{x} (\lambda_{2})_{y} (\lambda_{3})_{z} (P_1)^{m_1}_{x, x'} (P_2)^{m_2}_{y,y'} (P_3)^{m_3}_{z, z'} T^{\triangle, \triangledown}_{x' , y', z'},
\label{M-tensor}
\end{equation}
where $P_{1(2,3)}$ are three nearest-neighbor (NN) $P$-tensors, $m_1$, $m_2$, and $m_3$ represent physical indices, and $x$, $y$, and $z$ are geometric indices. Three-site projection operators $O_{\triangle, \triangledown} = \exp{(-\tau h_{\triangle, \triangledown})}$, for up (down) triangles, are projected on the $M$-tensor. We take successively such projections first simultaneously on all up triangles and then on all down ones till the convergence.

In each projection step, after absorbing the operator $O$ into $M$ and obtaining the evolved tensor $\tilde{M}^{m_1,m_2,m_3} = \sum_{m_1',m_2',m_3'} O_{m_1',m_2', m_3'}^{m_1,m_2,m_3} M^{m_1',m_2',m_3'}$, we need to decompose $\tilde{M}$ back into the product of $T$-, $P$-tensors and $\lambda$-vectors [inverse of \Eq{M-tensor}]. In this process, the bond dimension will increase and needs to be truncated. In order to achieve an optimal truncation, the environmental effects should be carefully taken into account. To be specific, we need to evaluate (exactly or approximately) the reduced density matrix $\rho^{x(y,z)}$ of the enlarged bond $x(y,z)$, which plays an important role in the truncation process.

There are basically two ways to evaluate the reduced density matrix. One way is to take exact contraction, i.e., we contract the reduced density matrix step by step from infinitely far away until it converges. Another (equivalent) way is to gauge the TTN tensors into their canonical form, and then the reduced density matrices could be obtained locally, which facilitates the evaluation of local observables. Here we use the second approach, and gradually bring the tensors into the canonical form (in an iterative way) during the course of imaginary-time evolution. A canonical TTN satisfies the following conditions simultaneously
\begin{eqnarray}
\sum_{m_1,m_2,m_3} \sum_{y,z} \tilde{M}_{x,y,z}^{m_1,m_2,m_3} \tilde{M}_{x',y,z}^{m_1,m_2,m_3} & = & \delta_{x,x'} \lambda_x \lambda_{x'},  \notag \\
\sum_{m_1,m_2,m_3} \sum_{z,x} \tilde{M}_{x,y,z}^{m_1,m_2,m_3} \tilde{M}_{x,y',z}^{m_1,m_2,m_3} & = & \delta_{y,y'} \lambda_y \lambda_{y'}, \notag \\
\sum_{m_1,m_2,m_3} \sum_{x,y} \tilde{M}_{x,y,z}^{m_1,m_2,m_3} \tilde{M}_{x,y,z'}^{m_1,m_2,m_3} & = & \delta_{z,z'} \lambda_z \lambda_{z'}.
\end{eqnarray}
More details about the canonicalization process can be found in  Ref. \onlinecite{Ran_Su}.

Given a canonical TTN, we can locally calculate the reduced density matrix of each bond and use its eigensystem to implement truncations. Take $x$-bond as an example, $\rho_{x,m_1; x', m_1'} = \sum_{m_2, m_3, y, z} \tilde{M}^{m_1, m_2, m_3}_{x, y, z} \tilde{M}^{m_1', m_2, m_3}_{x', y, z}$. Suppose that the eigenvectors and eigenvalues of three $\rho$'s are $U_{1 (2,3)}$ and $\Lambda_{1 (2,3)}$, we keep the largest $D_{c}$ eigenvalues and corresponding eigenvectors. The truncation operations are proceeded (on reshaped $\tilde{M}$-matrix) as
\begin{equation}
\tilde{T}_{\tilde{x},\tilde{y},\tilde{z}} = \sum_{x,y,z} \sum_{m_1,m_2,m_3} \tilde{M}_{x, y, z}^{m_1,m_2,m_3} (\frac{U_1}{\sqrt{\Lambda_1}})_{x, \tilde{x}}^{m_1} (\frac{U_2}{\sqrt{\Lambda_2}})_{y, \tilde{y}}^{m_2} (\frac{U_3}{\sqrt{\Lambda_3}})_{z, \tilde{z}}^{m_3},
\end{equation}
where $\tilde{x}$ ($\tilde{y}$, $\tilde{z}$) is the new bond index (truncated according to eigenvalues in $\Lambda$), and updated $\tilde{\lambda}_{1(2,3)} = \sqrt{\Lambda_{1(2,3)}}$ and $\tilde{P}_{1 (2,3)} = U^{-1}_{1(2,3)}$. This decimation scheme was introduced in Ref. \onlinecite{Ran_Su}, dubbed as the network Tucker decomposition.

In practical calculations, we set random or some fixed initial state (say, a dimer state) for the ground-state projections, and reduce Trotter slice $\tau$ gradually from $10^{-1}$ to $10^{-5}$ during the course of projections. For finite-temperature calculations, Trotter slice is usually chosen as $\tau=0.01$, and some extra loops (about 200) for gauging the TTN into a canonical form are necessary in every single evolution step.

\section{Quantum Spin Liquid Ground State}

\begin{figure}
\includegraphics[width=0.95\linewidth,clip]{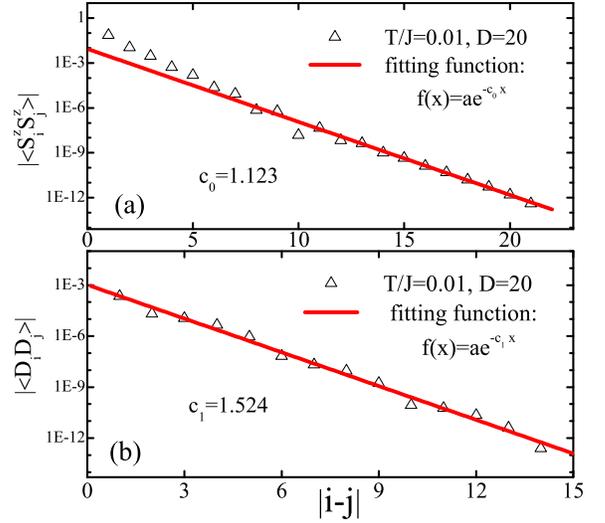}
\caption{(Color online) The spatial dependence of (a) spin-spin and (b) dimer-dimer correlation functions along the specified line ($\alpha$-$\beta$ line) in Fig. \ref{fig1}(a) on the infinite Husimi lattice, at low temperature ($T/J=0.01$). The fittings of the numerical data to an exponential function are presented for both correlation functions.}
\label{fig3}
\end{figure}

By performing the tensor-network based calculations, we studied the ground-state properties of the spin-1/2 XXZ HAF model on the infinite Husimi lattice, and unveiled that it is a gapless QSL. The unit cell of the Husimi lattice is a triangle consisting of three S=1/2 spins, thus the total spin of it can only be half-odd-integer. For anisotropy parameter $0 \leq \delta \leq 1$, the ground states are discovered to be non-magnetic, i.e., the local magnetic moment on any of the three sublattices vanishes. The spatial correlation functions of any local operators (say, spin operator $S_i^z$) are found to decay exponentially. In particular, the calculated energy per triangle for up- and down-triangle, and the bond energy for three different bonds in each triangle are the same within numerical errors. Therefore, we believe that this exotic ground state, with no spin rotational or lattice translational symmetry breaking, might be a featureless QSL.

In \Fig{fig2}, we present the results of energy per site $e_0$ and the correlation length $\xi$ against the bond dimension $D$ for $\delta=1$. The extrapolated energy $e_0=-0.4344(1)$ per site is very close to the best estimation of ground-state energy of kagome HAF model, $-0.4386(5)$, obtained by large-scale DMRG calculations.\cite{Yan, Depenbrock} The energy result confirms that it is this infinite Husimi lattice introduced here, instead of the finite-size Husimi cactus (with ground-state energy $-0.375$ per site), that could serve as a Bethe-lattice approximation of the counterpart model on a kagome lattice.

The correlation length $\xi$ = $-1/\ln[r(2)/r(1)]$, where $r(1)$ and $r(2)$ are the first- and second-largest eigenvalues of the transfer operator on the infinite Husimi lattice, is also plotted in \Fig{fig2}. $\xi$ is measured with the length unit of the underlying Bethe lattice,\cite{Li} which is shown to converge much more slowly than energy $e_0$ in \Fig{fig2}. Notably, owing to the special geometry of the Husimi lattice, a finite $\xi$ does not necessarily mean the existence of an excitation gap.

In \Fig{fig3}, we show the spatial dependence of the spin-spin correlation function $\langle S_{i}^{z} S_{j}^{z}\rangle$ along a path consisting of sites on $\alpha$ and $\beta$ sublattices [the red line in \Fig{fig1}(a)]. The spin-spin correlation is found to decay exponentially, as shown in \Fig{fig3}(a). Besides, we also calculated the dimer-dimer correlation function, defined as $\langle D_i D_j \rangle = \langle(S_{i}^{z} S_{i+1}^{z})\cdot (S_{j}^{z} S_{j+1}^{z})\rangle-\langle S_{i}^z S_{i+1}^{z}\rangle \cdot \langle S_{j}^{z}  S_{j+1}^{z}\rangle$, where the sites $i,j$ belong to the $\alpha$-$\beta$ line in \Fig{fig1}(a). The dimer-dimer correlator and its fitting are shown in \Fig{fig3}(b), revealing that the dimer-dimer correlation also decays exponentially. Other correlation functions, like the chiral correlation function $\langle C_{m} \, C_{n}\rangle=\langle [\overrightarrow{S}_{m_{1}}\cdot (\overrightarrow{S}_{m_{2}}\times \overrightarrow{S}_{m_{3}})] [\overrightarrow{S}_{n_{1}} \cdot (\overrightarrow{S}_{n_{2}}\times \overrightarrow{S}_{n_{3}})]\rangle$, where $m_1, m_2,m_3$ ($n_1, n_2, n_3$) denote the three sites in a triangle $m$($n$), are also calculated (not shown here). It is found that the chiral correlation is very weak even for a short distance, and it decays exponentially. Accordingly, the expectation value of the single loop operator
 $\langle C_{m}\rangle =
 \langle \overrightarrow{S}_{m_{1}} \cdot
 (\overrightarrow{S}_{m_{2}}\times\overrightarrow{S}_{m_{3}})
 \rangle$ is found to vanish, revealing the absence of a chiral order.

\section{Magnetization Curve and the Phase Diagram}

\begin{figure}
\includegraphics[width=0.9\linewidth,clip]{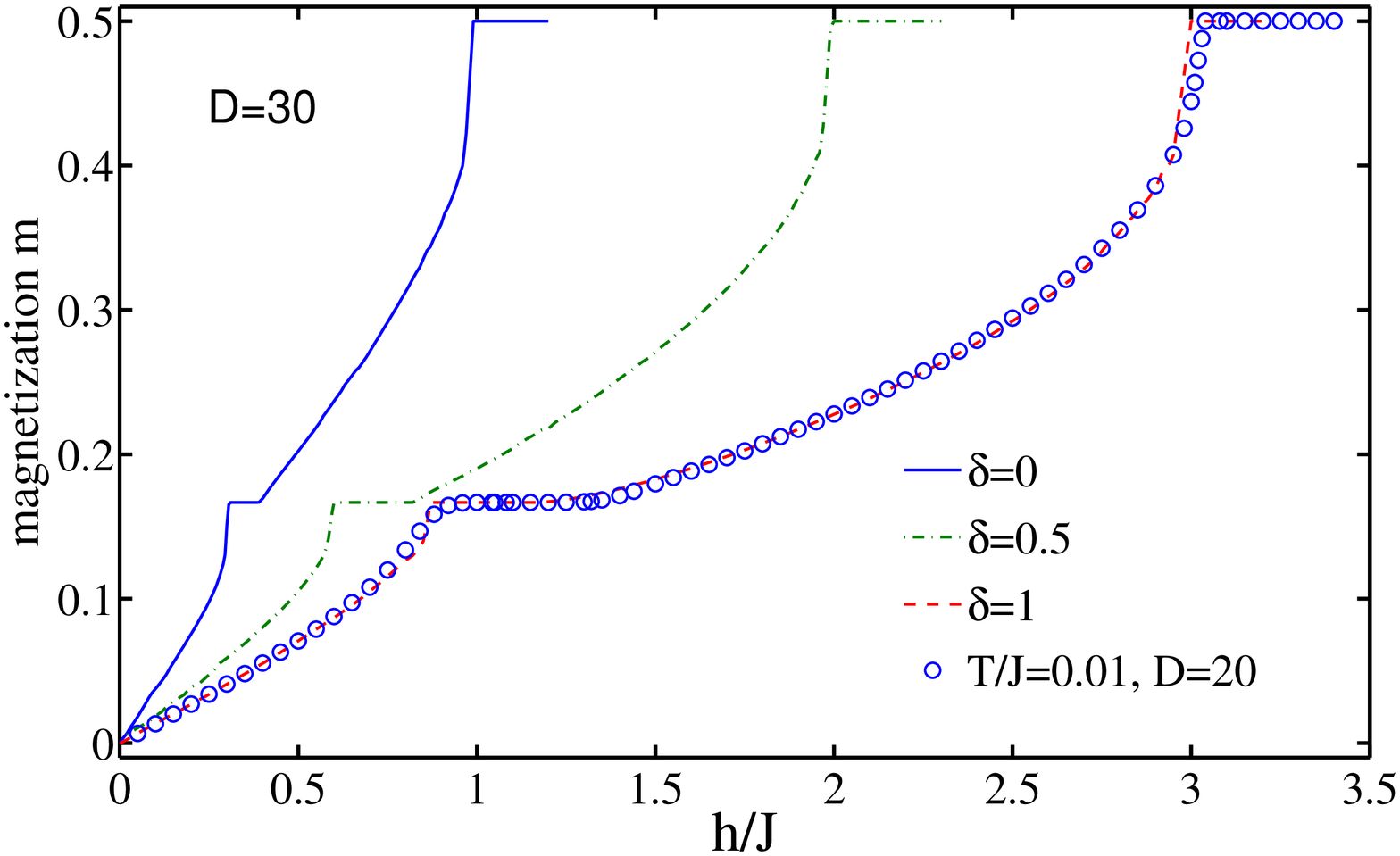}
\caption{(Color online) The magnetization curves of the XXZ HAF model ($\delta=0, 0.5, 1$) on the infinite Husimi lattice in the ground state. The 1/3-magnetization plateau exists for various $\delta$. The calculated magnetization curve at low temperature $T/J = 0.01$ is also presented for a comparison.}
\label{fig4}
\end{figure}

Next, we utilize jointly the projection approach for the ground state and the LTRG approach for thermodynamics to calculate the magnetization curves under uniform magnetic fields.
The ground-state magnetization curves for three typical cases ($\delta=0, 0.5, 1$) are plotted in \Fig{fig4}. For $\delta=1$, the LTRG results at very low temperature (T/J=0.01) are presented, showing a good accordance with the ground-state results. It is observed that for all these $\delta$, no matter the SU(2) isotropic Heisenberg model with $\delta=1$, or the XY model with $\delta=0$, the ground states are all non-magnetic when the magnetic field is absent, which can be attributed to the frustrations that enhance the quantum fluctuations and thus melt the spin orderings. Moreover, the zero-field susceptibilities are non-vanishing, and the local magnetizations $m(h)$ are linear (i.e., no zero-magnetization plateau) at small external fields $h$ for various $\delta$. This observation implies the absence of a spin gap.

Another interesting character in the magnetization curve is the apperance of a $1/3$-magnetization plateau, which has an intimate relation to the triangular motifs on the lattice. Similar plateaux have been observed in other materials or lattice models containing triangle motifs, e.g., triangular \cite{Ono, Alicea} and kagome\cite{Narumi, Schulenburg, Capponi} lattices. The 1/3-plateau in the former was explained with ``up-up-down" (UUD) spin structure on each triangle, and is therefore dubbed as an UUD phase, which has a quantum origin.\cite{Ono} To uncover the nature of this 1/3-plateau on the Husimi lattice, we compute the local magnetizations on three sublattices, revealing that it is indeed an UUD plateau on Husimi lattice. Interestingly enough, in \Fig{fig4}, we observe that this UUD 1/3-plateau is rather robust, which even exists for $\delta=0$. This remarkable observation manifests that the quantum fluctuation (XY-term in Eq. \ref{eq1}) selects and stabilizes this plateau state under certain fields.

It is an interesting issue to compare the magnetization curve with that of the kagome Heisenberg model. The unit cell of the Husimi lattice is a simplex consisting of three sites, hence only possesses one 1/3-plateau in the magnetization curve, while the unit cell of kagome lattice model varies with different magnetic fields. At zero field, the unit cell of the kagome lattice is also a simplex, while at 1/3, 5/9, and 7/9 magnetization plateaus it changes to a hexagram containing nine sites\cite{Capponi, Nishimoto}.

\begin{figure}
\includegraphics[width=0.9\linewidth,clip]{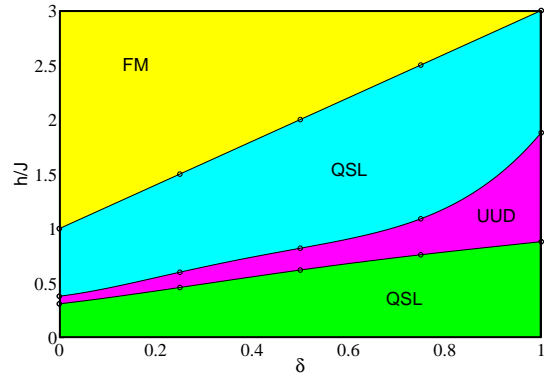}
\caption{(Color online) The $h - \delta$ phase diagram in the ground state of XXZ HAF model on the infinite Husimi lattice. FM stands for the field-induced ferromagnetic phase, QSL means quantum spin liquid (paramagnetic) phase, and the intermediate region labeled by UUD is the 1/3-magnetization plateau phase with up-up-down spin structure.}
\label{fig5}
\end{figure}

In \Fig{fig4}, we found that, for various $\delta$, there exist two continuous regions, one between $h=0$ and the lower critical field for the $1/3-$plateau, and the other between the upper critical field and the saturation one. These two regions own similar properties to a spin liquid state under zero field, except for that they have nonzero field-induced magnetizations along the $z$ direction. The spin liquid states behave like paramagnets in these regions. By summarizing the calculated results of magnetization curves for various $\delta$, we obtain a ground-state phase diagram in the plane of $h-\delta$ for the XXZ HAF model, as shown in \Fig{fig5}. It is seen that there are phases including the field-induced ferromagnetic phase, two QSL (paramagnetic) phases, and a 1/3-magnetization plateau (UUD) phase.

\section{Thermodynamic Properties}

Now we turn to explore the thermodynamic properties of the model by tensor-network algorithms following the same line developed in LTRG methods.\cite{LTRG, Ran_Su} The free energy can be obtained by collecting all the renormalization factors down to the particular low temperature that we set. The energy as well as other thermodynamic quantities can be obtained by taking derivatives of the free energy. Alternatively, we can also evaluate them by computing the expectation values of operators, like the local Hamiltonian, in the TTN thermal states.
\subsection{Susceptibility}
\begin{figure}
\includegraphics[width=0.95\linewidth,clip]{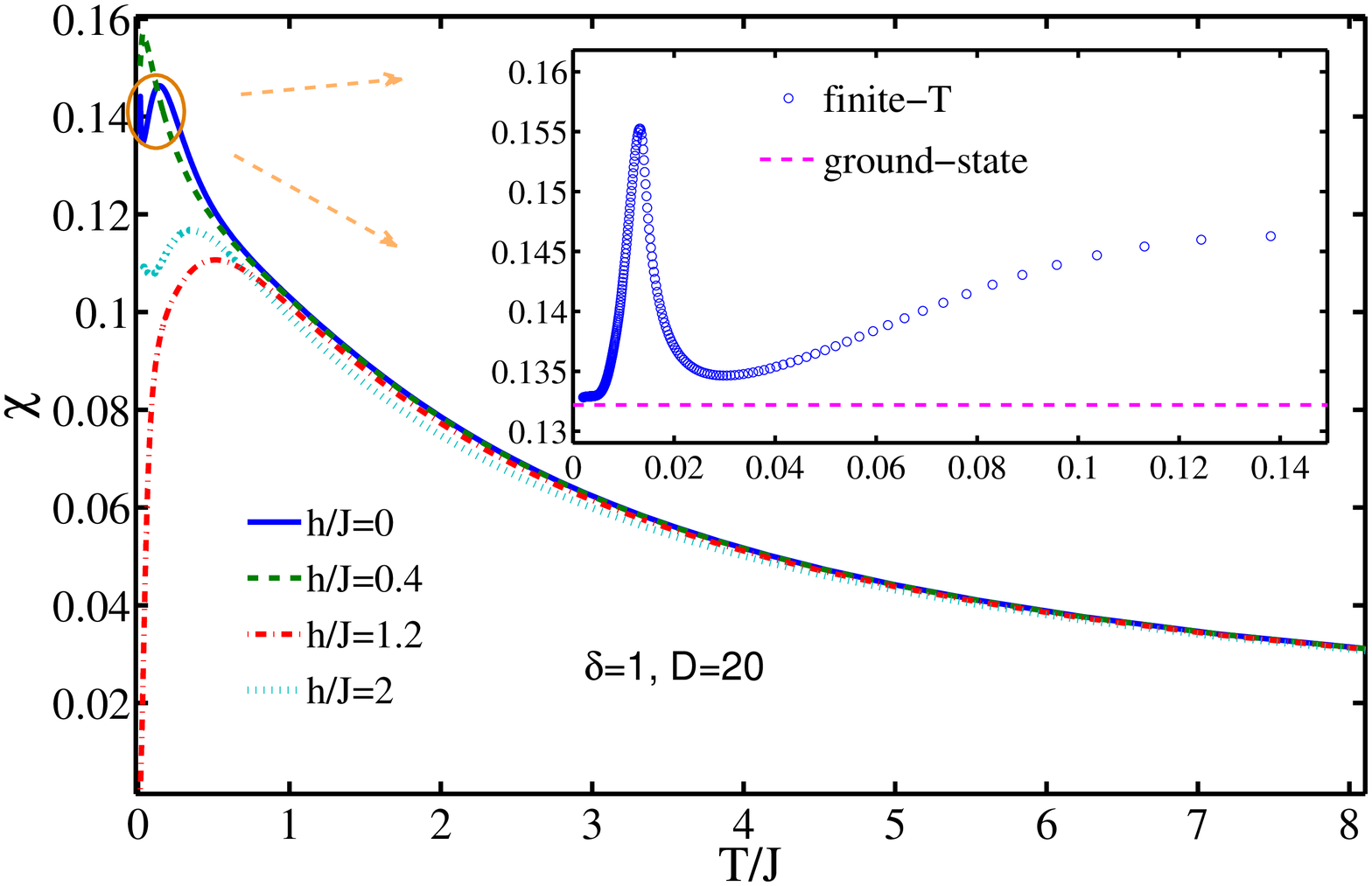}
\caption{(Color online) The uniform susceptibility $\chi = [m(h+\delta h) - m(h)]/\delta h$ as a function of temperature under various fields for the HAF model on the infinite Husimi lattice. $\delta h/J =0.02$ is taken. The inset shows two round peaks of $\chi$ at low temperatures in the spin liquid region, and the dashed line shows the ground-state susceptibility result (obtained by imaginary-time projections).}
\label{fig6}
\end{figure}

\begin{figure}
\includegraphics[width=0.9\linewidth,clip]{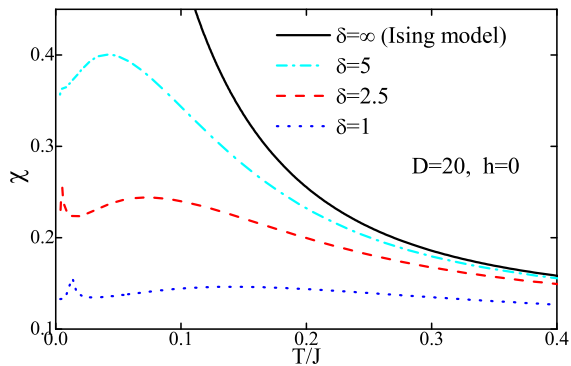}
\caption{(Color online) The susceptibility as a function of temperature for the XXZ HAF model on the infinite Husimi lattice in the absence of a magnetic field, where $\delta=1, 2.5, 5, \infty$. The susceptibility of Ising model is divergent in the zero temperature limit, while there exist two round peaks when the anisotropic parameter $\delta$ is finite.}
\label{fig7}
\end{figure}

The susceptibility $\chi$ as a function of temperature is plotted under various magnetic fields $h$, as shown in \Fig{fig6}. $\chi$ converges to different values at zero temperature, depending on which phase the system belongs to. In the spin liquid regime ($h/J=0, 0.4,$ or 2 in \Fig{fig6}), $\chi$ is nonzero at $T=0$, revealing the gapless feature of magnetic excitations; while for the UUD plateau phase ($h/J=1.2$), $\chi$ vanishes at zero temperature as expected, validating the existence of an excitation gap. Another impressive observation is the appearance of double peaks in susceptibility at low temperatures, which is scarce and peculiar for spin systems.

The bimodal structure of the susceptibility $\chi$ at low temperature is quite robust against varying the anisotropy $\delta$. In \Fig{fig7}, we show the susceptibility versus temperature for various $\delta$ under zero magnetic field. It is observed that for the bimodal structure of $\chi$, the left peak is quite sharp and depends weakly on $\delta$, while the right peak is broad and becomes more pronounced with increasing $\delta$. For the classical Ising limit ($\delta=\infty$), the susceptibility diverges at zero temperature. This low-temperature double-peak structure of susceptibility might be owing to the quantum fluctuations and geometric frustration effects.

\subsection{Specific heat}

\begin{figure}
\includegraphics[width=0.95\linewidth,clip]{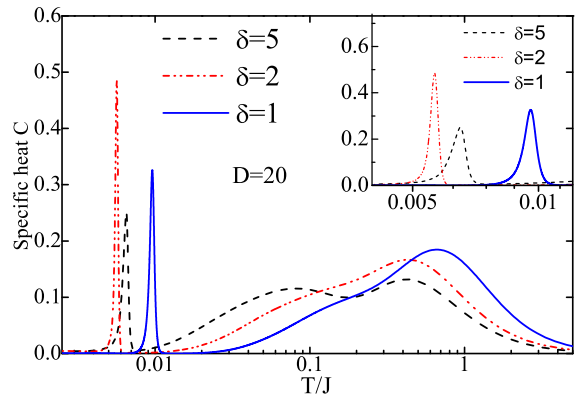}
\caption{(Color online) Temperature dependence of the specific heat of HAF model on the infinite Husimi lattice for various $\delta$ in the absence of a magnetic field. Three peaks are observed in each curve. Note the leftmost low-temperature peak, which is, though very sharp, a non-diverging round peak, as the magnified plot shows in the inset.}
\label{fig8}
\end{figure}

The overall landscape of the specific heat $C$ versus $T$ is quite complicated, as shown in \Fig{fig8}, which exposes at least two round peaks (for some cases, say, $\delta=5$, there are even three peaks), and  none of them are found to be divergent, reinforcing the statement that there is no symmetry breaking in the Husimi HAF model. Similar to the magnetic susceptibility curve, there exists a sharp (but not divergent) peak at very low temperature (the leftmost one in \Fig{fig8}), which is believed to have intimate relations with quantum fluctuations and geometric frustrations. In the absence of quantum fluctuations (classical Ising limit) or frustration effects (for instance, set the coupling on one of the three edges as ferromagnetic to eliminate the frustration), the leftmost low-T peak would disappear. Moreover, the position of this peak is found to be around $T/J = 0.005\sim0.01$, which changes slowly with $\delta$, as shown in the inset of \Fig{fig8}.

\begin{figure}
\includegraphics[width=0.95\linewidth,clip]{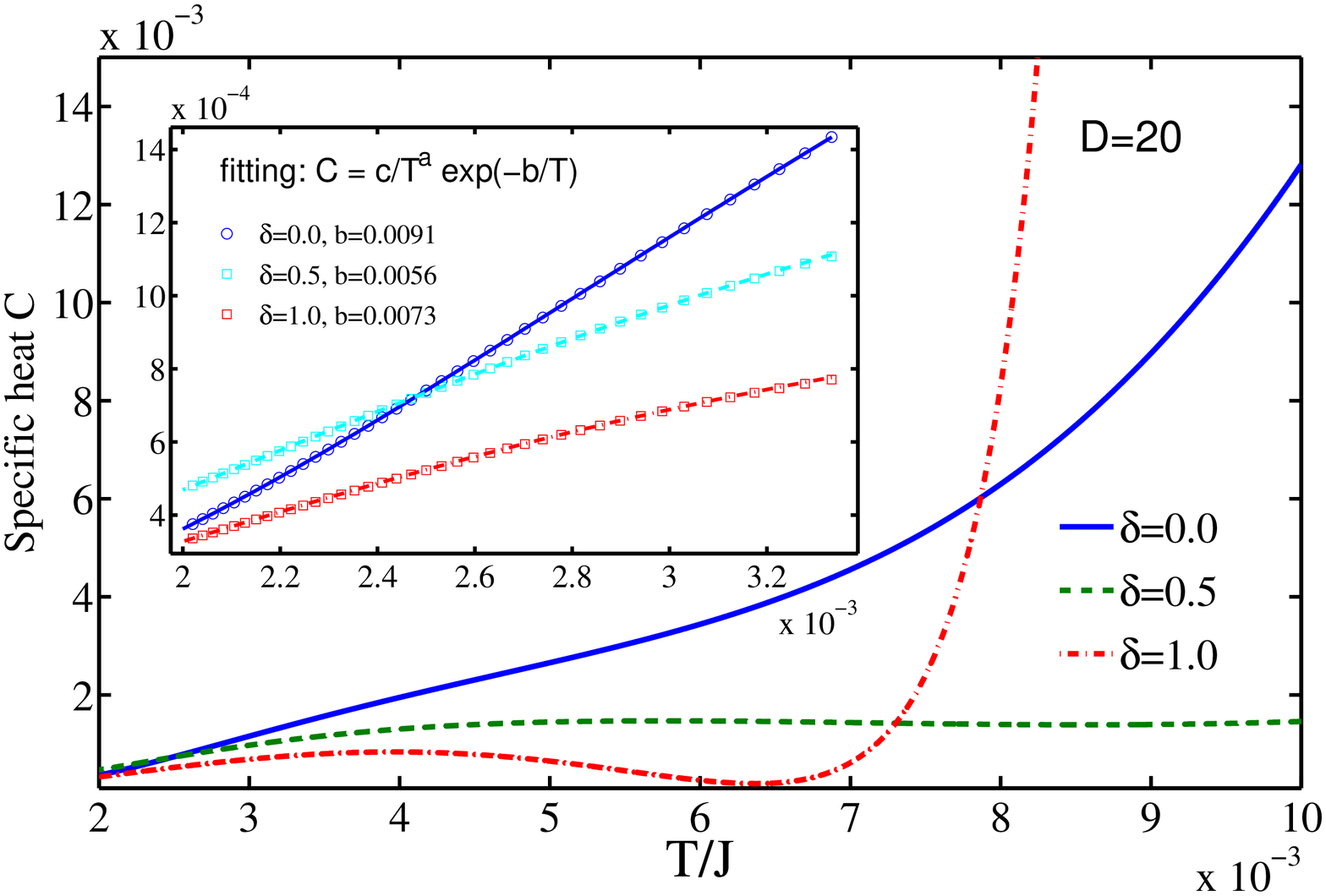}
\caption{(Color online)  The temperature dependence of the specific heat of the HAF model on the infinite Husimi lattice at extremely low temperature for various $\delta$, and in the absence of a magnetic field. The lowest temperature segments are amplified in the inset, and the fitting lines to an exponential decay are also included, which show that the excitation gap, if any, should be less than $10^{-2} J$ for all three cases.}
\label{fig9}
\end{figure}

In \Fig{fig9}, the temperature dependence of the specific heat $C$ at extremely low temperatures are presented for three cases with anisotropy $\delta=0, 0.5$ and 1. In the inset of \Fig{fig9} we amplify the very low temperature segment, which has almost linear $C-T$ relations. The fitting with an exponential decay of the form $C = \frac{c}{T^a}\exp{(-\frac{b}{T})}$ suggests that the gap (if any) is negligibly small up to the computational errors.

\section{Conclusion}
In this work, we study both the ground-state and thermodynamic properties of the spin-1/2 quantum HAF model on the infinite Husimi lattice. The ground-state is revealed to be a featureless disordered state without any spontaneous symmetry breaking, i.e, a quantum spin liquid state. The absence of the zero-magnetization plateau in the magnetization curves suggests that the spin excitation is gapless. A 1/3-magnetization plateau with up-up-down spin configuration is observed in the magnetization curve, which exists even when the spin-spin couplings are purely of XY-terms. The thermodynamic quantities including the specific heat and susceptibility are studied, and no signal of phase transition has been detected at any finite temperature. The algebraic decaying low-temperature specific heat, as well as the non-vanishing zero-field susceptibility, confirms the existence of a gapless and featureless quantum spin liquid.

\acknowledgments

We are indebted to B. Xi, H.-H. Tu, S. S. Gong, Z. C. Wang, Q. R. Zheng, and Z. G. Zhu for stimulating discussions, and  Z. Hao and O. Tchernyshyov for helpful communications. This work was supported in part by the NSFC (Grants No. 10934008, No. 90922033), the MOST of China (Grant No. 2012CB932900 and No. 2013CB933401), and the CAS. WL was also supported by the DFG through SFB-TR12 and NIM.

\end{document}